# Competing electronic ground states in the heavy-fermion superconductor CeRh$_2$As$_2$

Joanna Bławat[1*], Grzegorz Chajewski[2], Daniel Gnida[2], John Singleton[1], Oscar Ayala Valenzuela[1], Dariusz Kaczorowski[2], Ross D. McDonald[1*]

[1]National High Magnetic Field Laboratory, Los Alamos National Laboratory, Los Alamos, New Mexico 87545, USA

[2]Institute of Low Temperature and Structure Research, Polish Academy of Sciences, ul. Okólna 2, 50-422 Wrocław, Poland

## Abstract

CeRh$_2$As$_2$ is rare among superconductors, in that magnetic field tunes it between two distinct superconducting phases. Combined with a lack of local inversion symmetry and an upper critical field exceeding the Pauli paramagnetic limit, this excitingly suggests triplet multicomponent superconductivity. Preceding the superconducting onset, *f*-electron correlations cause long-range order, attributed both to local antiferromagnetism and itinerant (quadrupole) density-waves. Magnetic field provides a significant perturbation of the *f*-electron and may reveal the nature of the many-body correlations. Thus, we report comprehensive magnetization and magnetotransport studies on microstructured devices in fields of up to 73 T. Applied along the c-axis, field causes a low-temperature valence transition at $\mu_0 H \approx 24$ T. By contrast, in-plane fields produce a cascade of phase transitions; the field-induced in-plane conductivity anisotropy and lack of accompanying magnetic features, plus the closed-dome nature of the overall phase boundary is consistent with a hierarchy of field-induced density-wave states.

Corresponding authors: jblawat@lanl.gov; rmcd@lanl.gov

The electronic ground states in Ce-based heavy-fermion compounds are highly sensitive to the strength of coupling between *f*-electron orbitals and other bands, resulting in the formation of states ranging from long-range magnetic order to unconventional superconductivity or non-Fermi liquid behavior [1]. Furthermore, the energy differences between these states are frequently small, so that the compound's behavior is easily tuned by changes in composition, applied pressure and/or magnetic field [1–3]. As a result, such materials offer the possibility to study the interplay between competing orders in a single strongly correlated system. In addition to chemical doping and external pressure, high magnetic fields can induce valence transitions in Ce compounds. For example, fields induce either *f*-electron localization (e.g., elemental Ce, $CeOs_4Sb_{12}$ [4]) or delocalization ($CeCoIn_5$ [2]). Moreover, akin to the application of pressure, field can induce quantum critical behavior[4,5], nematicity[6,7], density waves[8], and unconventional superconductivity[3,9], all of which challenge current understanding.

Previous observations of multiphase superconductivity in heavy-fermion systems are generally restricted to Uranium compounds (e.g., $UTe_2$, $UPt_3$ [9–12]). However, recently $CeRh_2As_2$ was shown to exhibit multiple superconducting phases below $T_C = 0.26$ K [10]; magnetic field applied along the c-axis is thought to induce a transition from an even-parity to an odd-parity superconducting state at around 4 T [10]. The higher-field phase has an upper critical field of 14 T, exceeding the Pauli limit [10]. In addition, $CeRh_2As_2$ crystallizes in the centrosymmetric tetragonal P*4/nmm* space group [5], with Ce occupying two positions with opposite Rh and As environments, breaking local inversion symmetry. These properties led to predictions of multicomponent triplet superconductivity with a complex order parameter[13–17]. This fundamental materials research piqued the interest of the quantum-technology community; the bound states of topological superconductors are Majorana zero modes, leading to speculation that they are candidates for braiding quantum-computation schemes [14,18–21].

Preceding the onset of superconductivity, $CeRh_2As_2$ exhibits a second-order phase transition at $T_0 = 0.5$ K [22,23]. Initially, this was attributed to a quadrupole density wave (QDW)[10,24] of purely electronic origin; however, recent muon spin spectroscopy (μSR), nuclear magnetic resonance (NQR/NMR) and heat-capacity studies revealed that this transition also has a antiferromagnetic character. This may imply the coexistence below $T_C$ of magnetic order, QDW and superconductivity [22,25–27].

When the field is applied along the *a*-axis, only the low-field superconducting phase is observed, with an upper critical field of 2 T [10,17]. This strongly anisotropic response is reflected in the field dependence of the magnetic ordered phase which is suppressed by field applied along *c*, whilst it is stabilized to higher temperature by fields of up to 18 T [23,24] along the *a*-axis. Although subject to extensive low-temperature and low-field experimental and theoretical studies, the origin of and relationship between the multiple phases exhibited by $CeRh_2As_2$ remain elusive.

Motivated by these phenomena, we report high magnetic field studies on $CeRh_2As_2$ that follow the trajectories of transitions observed below 18 T and investigate the relationship between various competing/coexisting phases using magnetic field as a tuning parameter. We find a field-induced valence transition with **H** ∥ c, while the in-plane configuration exhibits a cascade of suggested density waves. By constructing the first high field *T-H* phase diagrams of $CeRh_2As_2$, we reveal the subtle competition between correlated ground states and the role of magnetic field, not only in tuning the electron-electron pairing responsible for different symmetry superconducting states, but also favoring electron-hole paring resulting in density-waves. Both valence transitions and density waves induced by magnetic field have been observed in $CeTIn_5$ (*T* = Ir, Rh, Co) suggesting that they might be a common feature among the heavy fermion unconventional superconductors.

**Field induced valence transition**

To help measure the magnetoresistivity tensor with current applied along specific crystallographic directions, and to increase the resistivity signal-to-noise ratio, Focused Ion Beam (FIB) lithography was employed to fabricate devices with lateral dimensions ~10 μm, resulting in 4-point resistances ranging from 100 mΩ to several ohms[28] with electric current flowing within the tetragonal a-b plane (*I* ⊥ c). To measure the resistance of $CeRh_2As_2$ with **H** ∥ c, an 8 μm thick rectangular Hall bar 340 μm x 40 μm was used [see Fig. 1(a), inset]. The temperature dependent resistivity down to 50 mK is presented in Fig. 1(a). It exhibits typical heavy-fermion behavior; i.e., first *R*(*T*) shows a shallow minimum on decreasing the temperature from 300 K, then forms a maximum at about 40 K, characteristic of the Kondo effect[1]. At low temperatures, there is a plateau below $T_0$, and a sharp drop to zero resistance at $T_c$ = 0.3 K related to the superconducting transitions. The critical temperature is in a good agreement with those previously reported[10,22,25]. The residual resistivity ratio RRR = *R*(300 K)/*R*(0.5 K) = 2, while in bulk $CeRh_2As_2$ single crystals the RRR was from 1.3 to 2.8 [25,29–32]. Investigations under fields of up to 14 T revealed the previously reported superconducting phases presented in Fig. 1(b). Our resulting phase diagram seems identical to that constructed earlier using transport, heat capacity, magnetization and magnetostriction data from bulk crystals[10,17,23,24,30]. Therefore, our measurements suggest that the FIB process does not unduly alter the intrinsic properties of $CeRh_2As_2$.

Fig. 2 (a) shows magnetoresistivity (MR = $[R(H) – R(0)]/R(0)$, where $R(0)$ is the zero-field resistance) measured up to 60 T with **H** ∥ [001] at temperatures between 0.6 and 80 K. Below 1 K, negative magnetoresistance is observed at low fields, followed by a rise to a maximum around 15 T. By contrast, for $1.2 < T < 4$ K the MR is small and positive up to $\mu_0 H \approx 20$ T, beyond which it drops drastically, reaching -60 % at 60 T. Although the high-field MR gets smaller with increasing temperature, it stays negative up to 80 K. Fig. 2(b) presents the low temperature Hall resistivity $\rho_{xy}$ versus field. Throughout the measured temperature range $\rho_{xy}$ remains positive, indicating that the dominant charge carriers are holes (see Fig. S1). At 0.65 K, $\rho_{xy}$ is linear in field up to $\mu_0 H \approx 24$ T, beyond which it shows a marked decrease. Above $\mu_0 H \approx 40$ T, $\rho_{xy}$ again becomes linear, but with a slope approximately half that at low fields. The red lines in Fig. 2(b) are linear fits through $H = 0$ emphasizing the change in slope of $\rho_{xy}$ between the low- and high field regions. The net carrier concentration $n = (n_h-n_e)$ ($n_e$ and $n_h$ are electron and hole densities respectively) extracted from the Hall coefficient via $R_H = 1/ne$ is $n \approx 5.16 \times 10^{21}$ cm$^{-3}$ at low fields and $n \approx 9.92 \times 10^{21}$ cm$^{-3}$ at high fields. Blue lines in Fig. 2(b) indicate the expected slope of $\rho_{xy}$ for exactly 1 and 2 holes/formula unit, showing that the transition observed in the experimental $\rho_{xy}$ data corresponds to an increase of almost exactly one hole per unit cell (volume 181.14 Å$^3$).

Because metals with relatively high carrier concentrations exhibit a tiny Hall signal, $\rho_{xx}$ has a significantly better signal-to-noise ratio [Fig. S1]. Consequently, to examine how the transition to increased hole density evolves with temperature and applied field we calculated the derivative $dR_{xx}/dH_z$ from the resistance $R(H)$ measured at various temperatures (Fig. S1). The derivatives exhibit two main features: (1) a maximum at $\mu_0 H \approx 15$ T, which is observed only at 0.68 K and 1.11 K; and (2) a broad minimum at $\mu_0 H \approx 24$ T and 0.68 K (i.e., at the same field as the change in $\rho_{xy}$ behavior) which moves towards higher fields with increasing temperature. Following the minimum in $dR/dH$ to higher $T$ results in the ($H, T$) phase boundary for **H** ∥ [001] shown in Fig. 2(c).

Taken together, the significant drop in the resistance and the low and high field slopes of $\rho_{xy}$ corresponding to an increase of carrier concentration suggest that the boundary in Fig. 2(c) is attributable to some form of field-induced valence transition, perhaps similar to that observed in $CeOs_4Sb_{12}$[4]; valence transitions are also known to occur in several other heavy-fermion materials (e.g., $YbInCu_4$, $URu_2Si_2$, elemental Ce, Pu [33–37]), resulting in a variety of phenomena. In the current context, such a valence transition may represent the field driving $CeRh_2As_2$ from a phase with a high Kondo temperature (and therefore *spd* and *f* hybridization, resulting in itinerant *f*-electron quasiparticles) to one with a low Kondo temperature in which the *f* electrons are in effect localized on the 4*f* multiplet [34,36]. A fall in chemical potential results, increasing the number of holes, observed both here and in $CeOs_4Sb_{12}$ [33]. In contrast to valence transitions in pure Ce and its alloys[36], the initial slope of the (*H,T*) phase boundary in Fig. 2(c) is positive. An analogous effect occurs in $CeOs_4Sb_{12}$ [4], attributed to proximity to a quantum-critical point associated with field-induced destruction of an AFM spin-density-wave (SDW) present below 1 K. The situation in $CeRh_2As_2$ may be closely related; the properties of its low-field, low-temperature (<0.5 K) AFM state seen in μSR and NMR experiments are consistent with a SDW [25,27].

Another possible contribution to (or explanation for) the positive slope of the phase boundary in Fig. 2(c) may come from predictions of Field Induced Quadrupolar ordering caused by coupling between the cerium $\Gamma_7$ ground state doublet and $\Gamma_6$; it results in an effectively localized quartet[38] at high fields. The latter effect may enhance the tendencies of the 4*f* electrons to quasi-localization, or even be functionally equivalent to a valence transition in removing the *f* electrons from the ensemble of itinerant carriers; the predicted phase boundary is qualitatively similar to Fig. 2(c).

To further elucidate the nature of this transition, we performed magnetization measurements (Fig. 3). For field along the primary crystallographic axes (**H** ∥ [001], [100], [110]), the magnetization is largely featureless, suggesting an absence of metamagnetic transitions and no significant changes in net magnetic moment; this is consistent with valence transitions in Ce alloys and compounds, which affects the conductivity but leads to relatively weak changes in the magnetization[33,36]. The magnetization is smaller when **H** ∥ [001] than with field applied in-plane, indicating that *c* is the hard axis up to 60 T. When the field is applied in the in-plane directions, the magnetization as a function of field looks almost identical at 0.62 K, but it decreases with temperature faster when **H** ∥ [100] than with **H** ∥ [110]. The lack of low-field in-plane anisotropy is consistent with the P4/*nmm* structure, with strong fields inducing a weak four-fold magnetic anisotropy.

**Cascade of in plane field induced density waves**

To further explore this in-plane anisotropy, structures were fabricated to enable control of both the in-plane field and current directions. Fig. S2(a, inset) shows a 20 μm thick device patterned into a V-shape to enable simultaneous measurement of the resistance along the [100] and [110] directions. The magnetoresistance as a function of field applied parallel to the current at various temperatures is plotted in Figs. S2(b) and (c), showing multiple features and a strong in-plane anisotropy of the conductivity. Devices with a longer meandering current path were also prepared to increase the resistance whilst maintaining control of the predominant current direction (both parallel to [100] and [110]) to enable higher-fidelity measurement of these subtle changes in conductivity in pulsed fields; these devices are in Figs. 4(a) and (b). The resistance as a function of field for a range of temperatures with **H** ∥ *I* is shown in Figs. 4(c) and (d). The overall behavior of $R(H)$ is similar to that observed in the V-shaped device, but the geometry enhances the resistance by an order of magnitude, indicating robustness of the observed in-plane field dependence (i.e., little or no sample to sample variation).

With **H** ∥ *I* ∥ [100], at 0.5 K, $R(H)$ decreases with increasing field, showing a minimum around 8 T, followed by an increase, with a broad maximum around 15 T. At higher fields $R(H)$ increases, exhibiting a maximum around 40 T before dropping quasilinearly to 60 T without any sign of saturation. With field and current applied along [110] at 0.5 K, the resistance similarly decreases with increasing field, with a local maximum around 8 T, but is followed by almost constant $R(H)$ from 10 to 28 T. Around 30 T, there is a relatively sharp increase in $R(H)$ with a very broad maximum up to 40 T, whence $R(H)$ decreases linearly to 60 T. As temperature increases, all features move to higher fields.

To assess whether the changes in magnetoresistivity are sensitive to both the applied magnetic field and current direction, measurements were repeated with field perpendicular to current, **H** ⊥ *I*, i.e. *I* ∥ [100] and **H** ∥ [010] (Fig. 4(e)), and *I* ∥ [110] and **H** ∥ [1$\bar{1}$0] (Fig. 4(f)). In the first case, $R(H)$ drops with increasing field, showing a change in slope around 8 T followed by a broad minimum above 20 T. After that, $R(H)$ increases, with a change in slope around 40 T. In Fig.

4(f), $R(H)$ shows a maximum around 8 T, followed by almost constant $R(H)$ from 10 to 30 T. At the latter field, a drastic increase in resistance is observed, followed by a very sharp decrease at 40 T and negative magnetoresistance to 60 T. Note also that significant hysteresis is observed between 30 and 40 T (Fig. S3), suggesting a first-order phase transition. For a selection of temperatures, these measurements were extended up to 73 T (Fig. S4) revealing an additional feature that moves to higher field on lowering $T$. Following the methodology employed for the $c$-axis MR, phase boundaries are ascribed to local minima/maxima in $dR/dH$ (Figs. S5, S6).

For each field direction ([100], [110]), despite the form and magnitude of the magnetoresistance depending upon the orientation of the current (parallel or perpendicular to field), within the experimental error the phase boundaries appear solely determined by the magnetic field orientation. The $T$-$H$ field phase diagrams with a typical resistance curve are presented in Fig. 5. Though not identical, the phase diagrams for **H** ∥ [010] and [110] are similar. Below 4 K, three closed field induced phases are observed, which evolve differently with increasing temperature.

Despite not being able to measure the Hall effect for in-plane-field orientations the changes in MR associated with the phase boundaries are not consistent with the large change in carrier density associated with the valence transition seen with field parallel to [001]. Furthermore, given the "closed dome" nature of each phase in the $T$-$H$ plane (solely determined by field orientation), the lack of observed change in magnetic susceptibility upon traversing the phases, and the field-induced anisotropy of the magnetoresistance in the in-plane phases are most consistent with a hierarchy of field-induced electronic density waves[39].

The calculated Fermi surface of $CeRh_2As_2$ [40–42] comprises multiple quasi-2 and 3-dimensional (D) pockets that are strongly dependent upon the strength of hybridization and $f$-electron valence. However, a common feature appears to be the propensity for planar sections of Fermi surface parallel to both [100] and [110] to appear at different heights ($k_c$) within the Brillouin zone. This suggests that field tuning of a density-wave nesting vector along [100] and [110] crystallographic directions is a likely explanation for the variation between in-plane phase boundaries. Density waves are both stabilized by magnetic fields owing to the effective reduction of dimensionality due to cyclotron motion and ultimately destabilized by Zeeman splitting, analogous to the Pauli paramagnetic limit of superconductors [43–45]. Comparing the maximum transition temperature (≈3.8 K) to the field that ultimately suppresses order (70 T) suggests that these density wave states are in the strong coupling limit[39].

Although these transitions appear to be predominantly electronic in nature, given that $CeRh_2As_2$ is on the cusp of a *f*-electron valence transition, it is interesting to speculate as to whether the observed phases involve purely charge, spin, or higher order density waves coupling multiple electronic degrees of freedom. According to Zanchi et al. [43], in quasi-1D systems with CDWs, the nesting vector does not get destroyed by applied magnetic field; instead unconventional DWs may be induced due to both charge- and spin-density-wave correlations [43]. $CeRhIn_5$ is not a 1D material; however, similar phenomena are proposed within a complex Fermi surface where both quasi-1D or 2D Fermi sheets coexist with 3D sheets [8]. The features observed in $CeRh_2As_2$ are remarkably similar to the cascade of field-induced charge density waves in $(Per)_2Pt(mnt)_2$ [44–46], graphite [47–49], and $CeRhIn_5$ [8]. When this is considered alongside the effect of magnetic field upon its superconducting phases, it becomes clear that $CeRh_2As_2$ is an important single material system for experimentally and theoretically investigating competition between electron-electron pairing, resulting in different symmetry superconducting states, and electron-hole paring, resulting in density-waves.

In this work, we present the first high field *T-H* phase diagram of $CeRh_2As_2$, and reveal the competition between correlated ground states and the role of magnetic field. Interestingly, there are phases of a different nature depending on the orientation of applied magnetic field. For **H** ∥ [001], there is an unusual field-induced valence transition, whilst in-plane magnetic fields (**H** ∥ [100] and [110]) induce a cascade of density waves. This suggests that magnetic field not only tunes the electron-electron pairing resulting in different symmetry superconducting ground states but also promotes electron-hole paring, resulting in density-wave ground states. As proposed in Ref. 22, many of the low-field properties of $CeRh_2As_2$ seem to be analogous to those of $CeRhIn_5$; our high-field studies further support those findings. Both valence transitions and density waves induced by magnetic field have been observed within other heavy fermion unconventional superconductors and metals, perhaps suggesting overarching physics common to these systems.

## Methods

### Single crystal synthesis and sample quality

Single crystals of $CeRh_2As_2$ were grown using the horizontal flux-growth technique. The detailed procedure is described in Ref. [30] Crystal orientation was checked using a PROTO COS Laue diffraction camera. Low-field resistivity measurements were performed using a Quantum Design Physical Property Measurement System (PPMS) equipped with a $^3$He-$^4$He dilution refrigerator. These data were used to create the low-field – low-temperature phase diagram presented in Figure 1(b). Points were selected using values (1) from temperature sweeps at different constant magnetic fields at 90 % and 50 % of the resistance drop, and where the resistance reaches the zero value (blue points), and (2) magnetic field sweeps at selected constant temperatures by choosing the local minima $H_1$ and $H_2$ in the magnetoresistance. The obtained phase diagram looks identical to that reported before constructed using transport, heat capacity, magnetization and magnetostriction measurements[10,17,23,24,30]. These low-temperature and low-field data were essential to confirm the properties of the FIB-fabricated sample in the range where the phase diagram is already well known and established by various thermodynamic probes; hence the samples can be studied with confidence in high magnetic fields, where the physical properties remained unknown until this work.

### Pulsed field magnetization measurements

The high field measurements were performed at the National High Magnetic Field Laboratory Pulsed-Field Facility at Los Alamos. The magnetization was measured using a compensated coil extraction magnetometer [50]. The crystal was placed in a 1 mm diameter ampoule which then can be moved in and out of the coil. The signal induced in the coil is proportional to the rate of change of magnetization with time d$M$/d$t$; the magnetization data are obtained by subtracting the integrated data without the sample present from that with the sample in the coil.

### FIB devices fabrication and high field magnetotransport measurements

The resistivity and Hall effect data were measured on a microstructured devices fabricated using a Focused Ion Beam (FIB) FEI Helios Nanolab 600. The single crystal of $CeRh_2As_2$ was hand-polished to 8-20 μm thickness. Transport and Hall measurements are extremely challenging in pulsed fields. Thinning down the sample and FIB fabrication made the experiment possible;

however the signal to noise is significantly lower in the Hall signal than that observed in the $R(H)$ curves. The noise values approach 10nV/√Hz, close to the limit for room temperature amplification at frequencies of 100's of kHz. Parts of the sample were sputter-coated with gold for electrical contacts, which were made using 25 μm Au wire and silver epoxy. At the final stage, the samples were microfabricated using FIB. In order to minimize the noise during the pulsed field measurements, each Au wire was attached to 50-gauge insulated copper wire twisted in current ($I$), voltage ($V_{xx}$) and Hall voltage ($V_{xy}$) pairs. For all measurement configurations a range of currents were applied without the observation of a non-linear IV response; this was checked in both the directly digitized sample voltage sine wave and the harmonic content of the digital lock-in output, indicating that all reported data were measured in the Ohmic response regime.

**Acknowledgments**

We would like to thank Philip Moll for constructive and helpful discussion. The work was performed at the National High Magnetic Field Laboratory, which is supported by National Science Foundation Cooperative Agreement No. DMR-2128556 and the State of Florida. JB and RM are supported by the Center for the Advancement of Topological Semimetals, an Energy Frontier Research Center funded by the U.S. Department of Energy, Office of Science through the Ames National Laboratory under contract DE-AC02-07CH11358. JS acknowledges the DOE BES FWP "Science of 100 T" that permitted the development of experimental techniques used in this paper, and which supports measurements above 65 T.

**Figures**

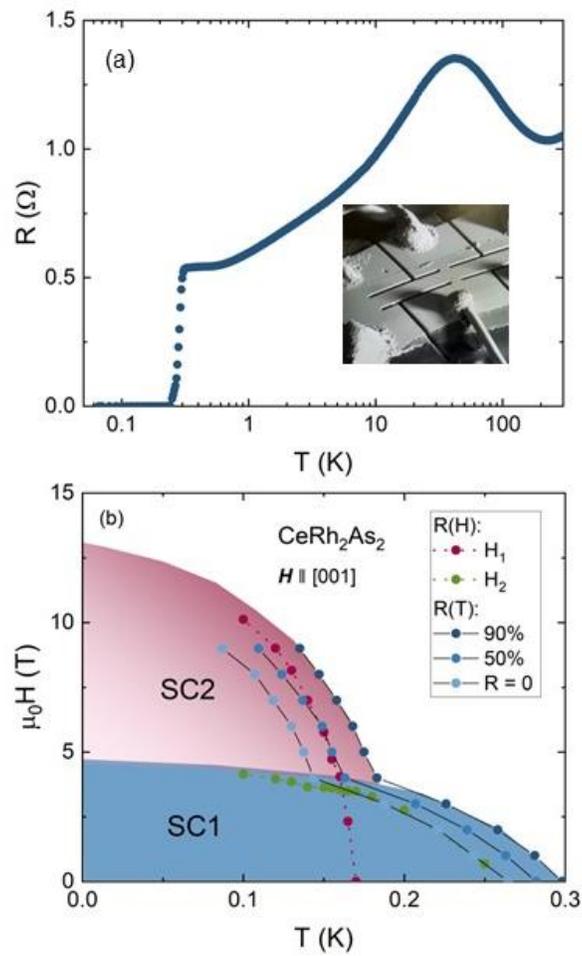

**Fig. 1. Low-field properties of FIB device and multiphase superconductivity.** (a) Resistance as a function of temperature measured on the FIB fabricated device. Inset) Electron micrograph of the Focused Ion Beam defined Hall bar structure used for this measurement. (b) low temperature phase diagram with **H** ∥ [001] created based on resistivity measurements on the FIB device showing two-phase superconductivity.

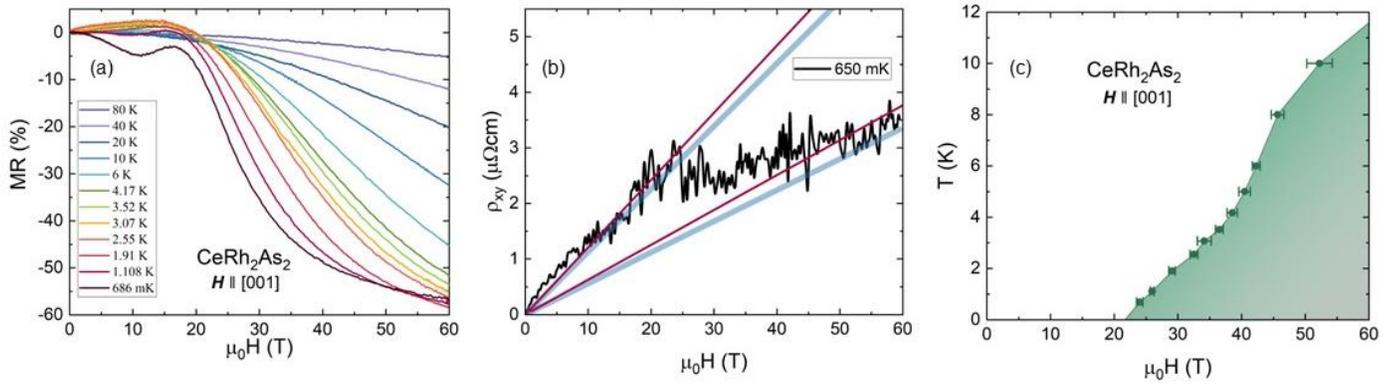

**Fig. 2**. **Field induced valence transition.** (a) Magnetoresistance of CeRh$_2$As$_2$ measured up to 60T for a range of temperatures between 600mK and 80 K. (b) Hall effect as a function of magnetic field. Red and blue lines correspond to line fit discussed in the text (c) Magnetic field – temperature points bounding the region of valence transition as determined by description in the main text, the error bars are determined from a parabolic fit to the minimum of dR/dH.

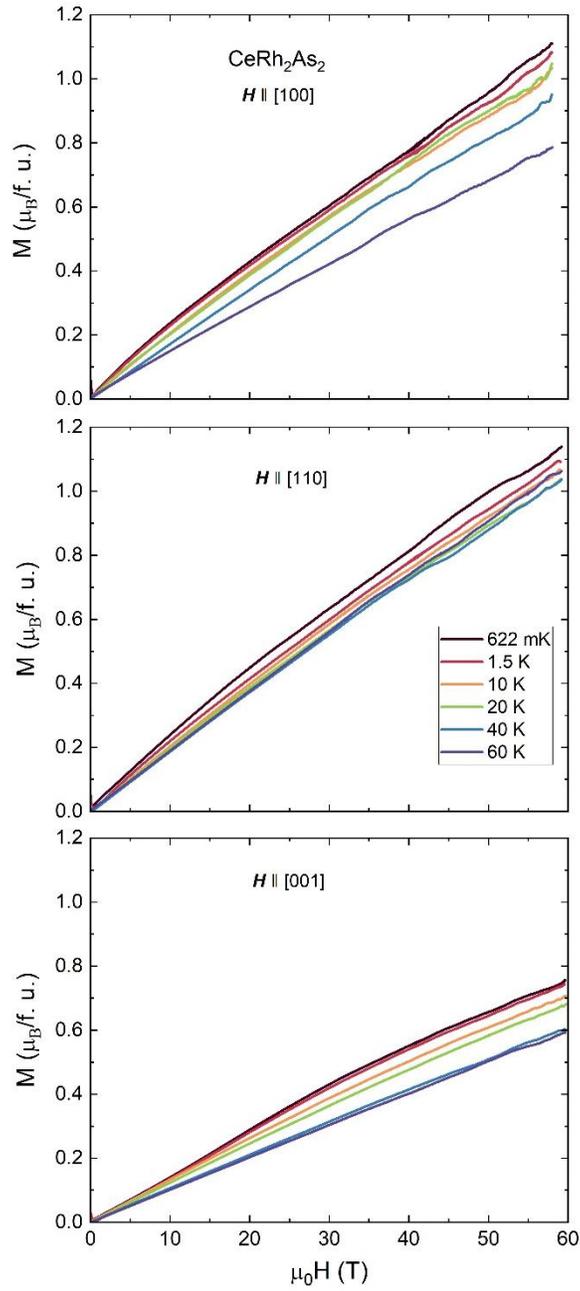

**Fig. 3. The featureless magnetization.** Magnetization of CeRh$_2$As$_2$ measured up to 60 T with magnetic field along the [100], [110], and [001] crystallographic directions.

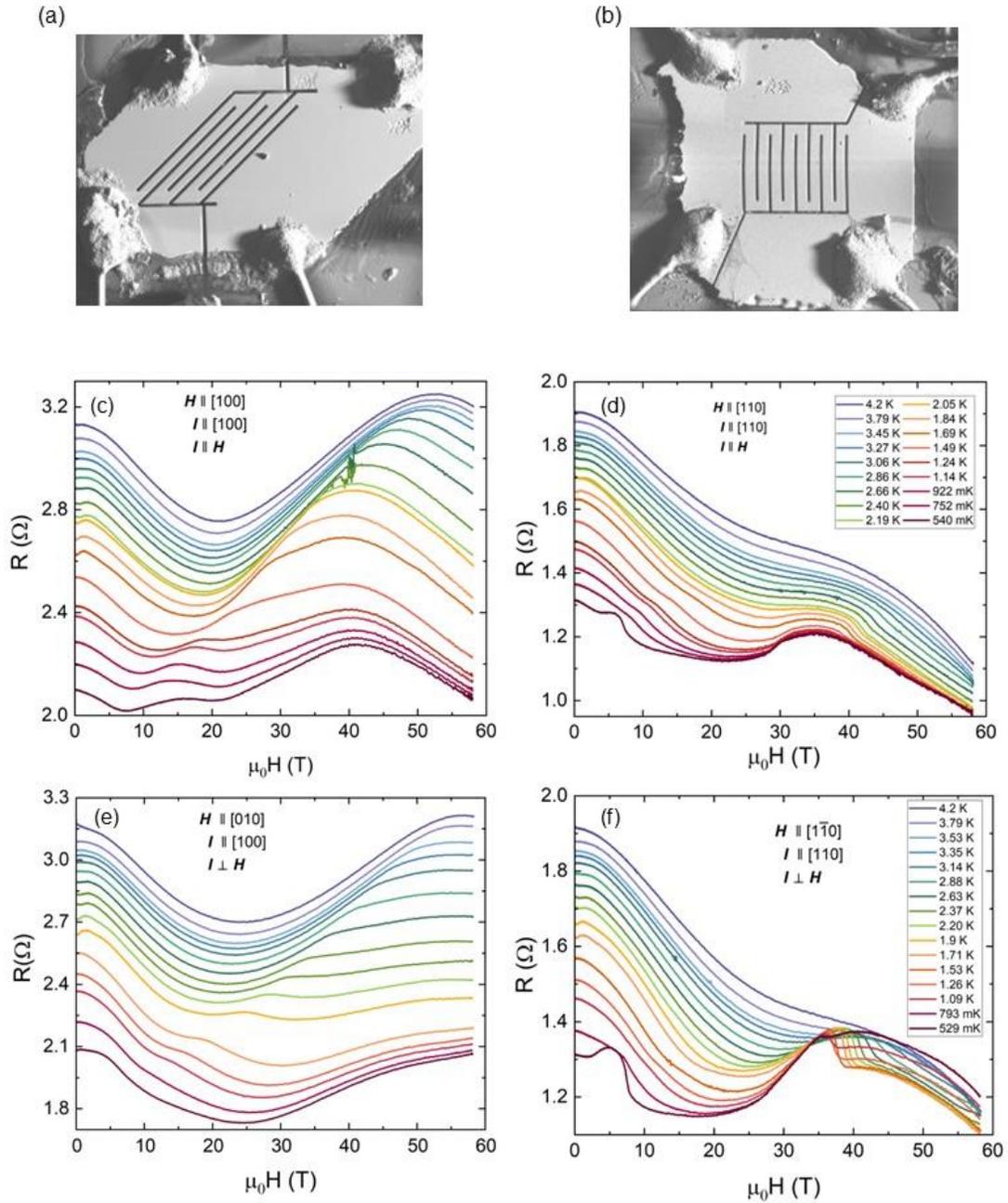

**Fig. 4. In-plane magnetoresistance on FIB microstructured devices.** Electron micrograph of the Focused Ion Beam devices with spring – like meanders microstroctured along [100] (a), and [110] (b). Resistance versus magnetic field with **H** ∥ **I** ∥ [100] (c), **H** ∥ **I** ∥ [110] (d), **I** ∥ [100] and **H** ∥ [010] (e), **I** ∥ [110] and **H** ∥ [1$\bar{1}$0] (f).

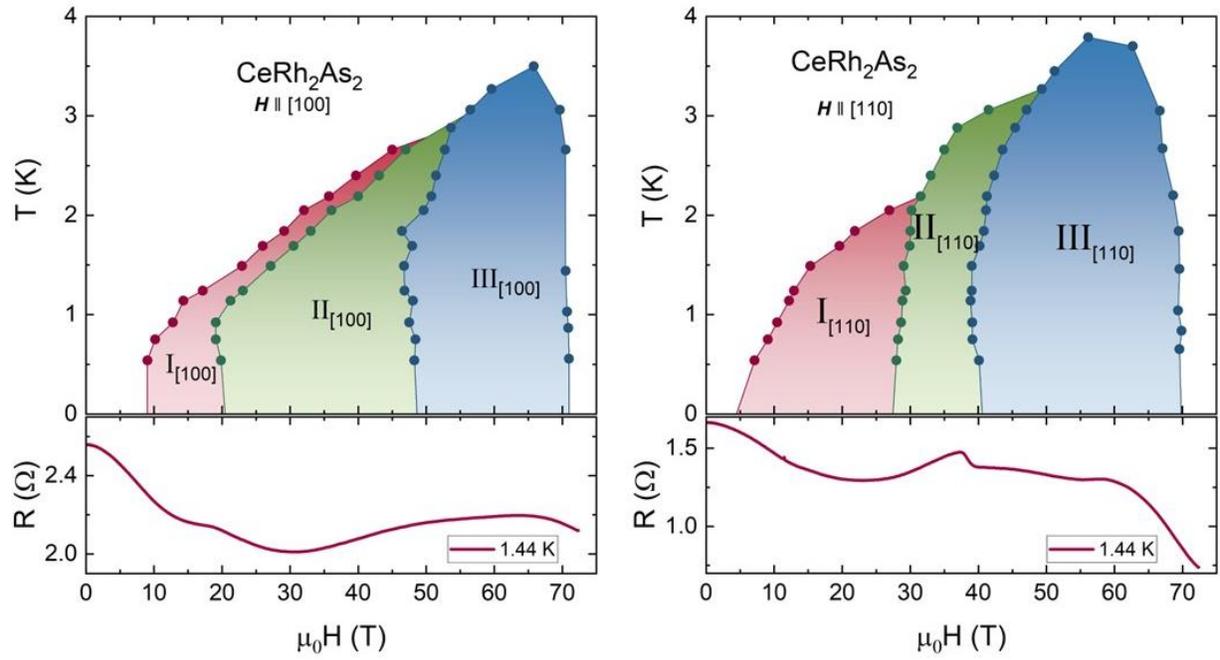

**Fig. 5. High field *T-H* in-plane phase diagrams of CeRh$_2$As$_2$** Temperature – Magnetic field phase diagrams for **H** ∥ [100] and **H** ∥ [110] with representative resistance versus magnetic field curve measured at 1.44 K up to 73 T.

# Supplemental Materials

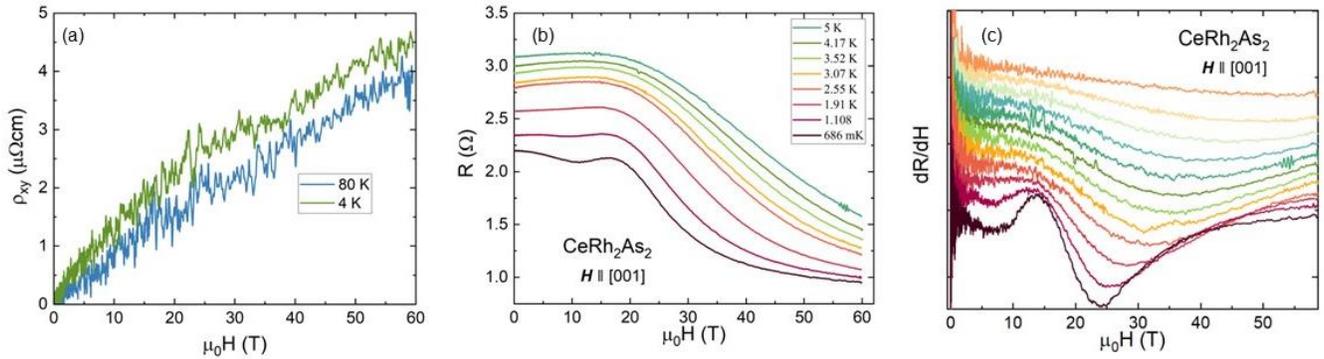

**Fig. S1.** (a) Hall resistivity as a function of magnetic field. (b) Resistance as a function of magnetic field measured at various temperatures with H ∥ [001] with corresponding derivative dR/dH (c) The derivative dR/dH as function of magnetic field measured with magnetic field along [001]

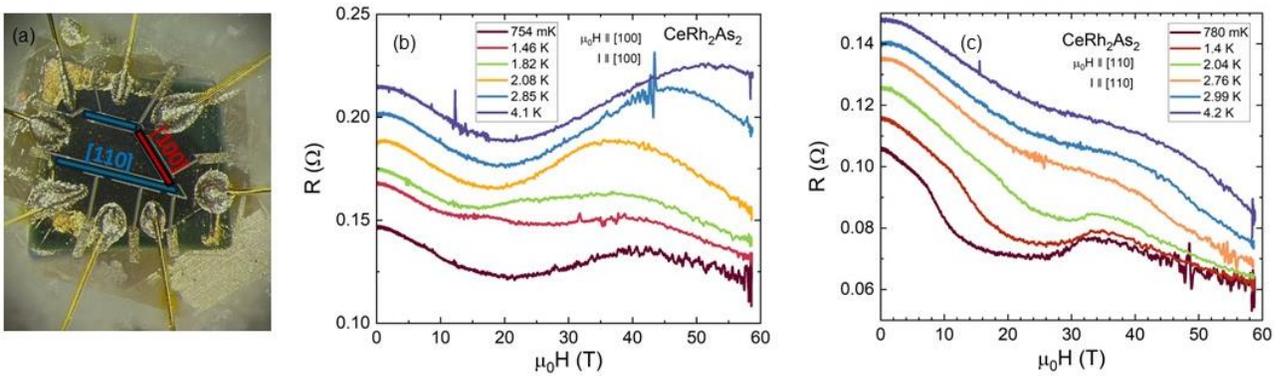

**Fig. S2.** Microscopic picture of the V-shaped FIB fabricated device with current flowing along [110] and [100] direction. Resistance versus magnetic field at various temperatures with H ∥ [100] (b) and H ∥ [110] (c).

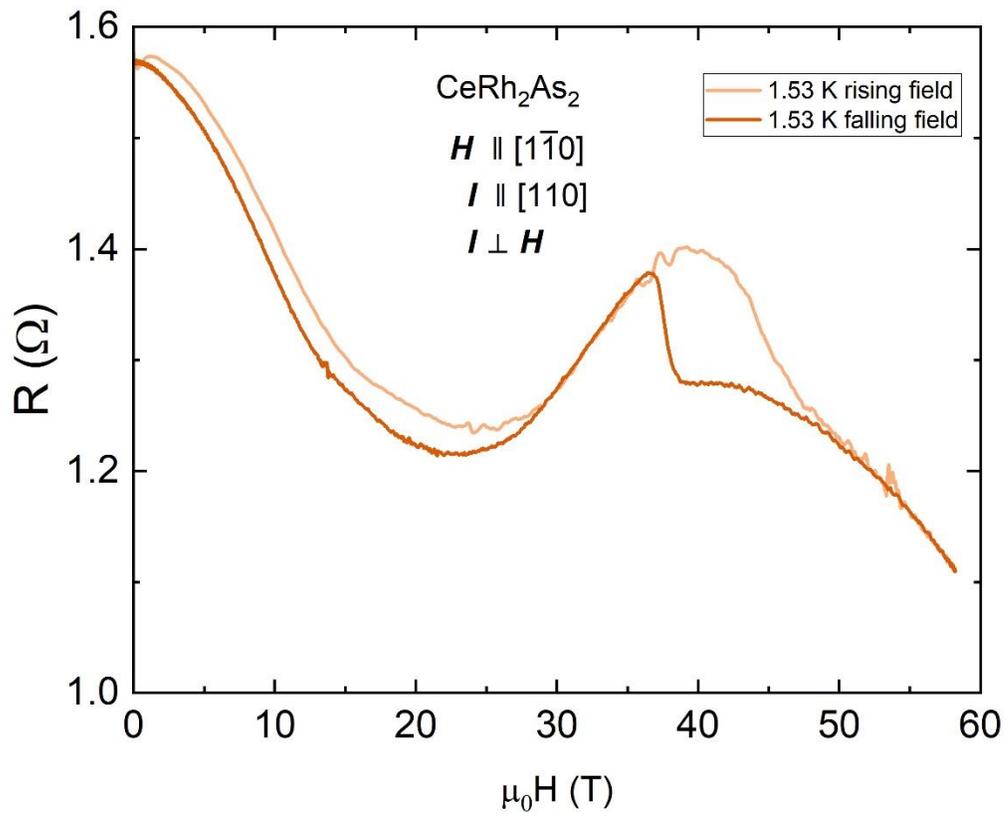

**Fig. S3.** Resistance as a function of magnetic field measured at 1.53 K showing the hysteretic behavior of the field induced transition.

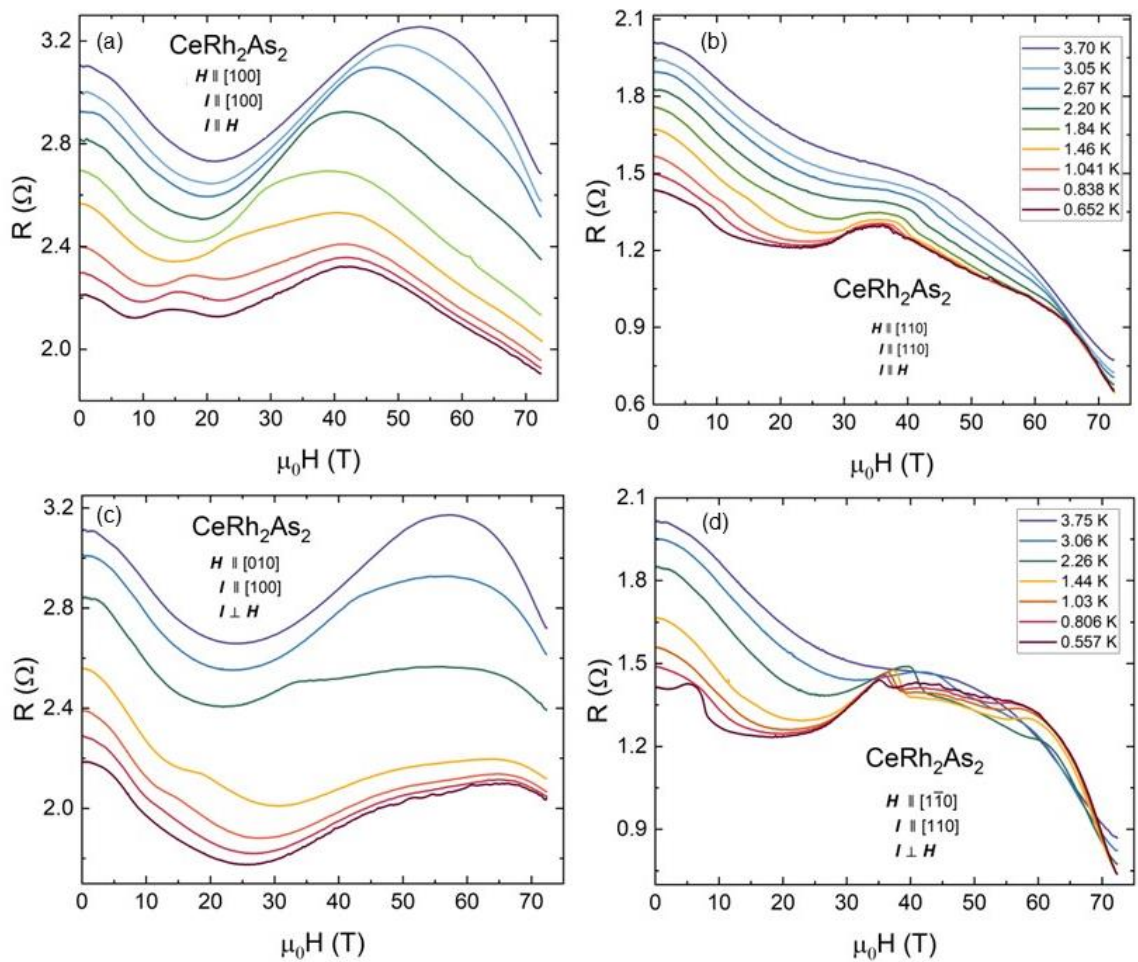

**Fig. S4.** The resistance data measured in the duplex magnet in magnetic field up to 73 T.

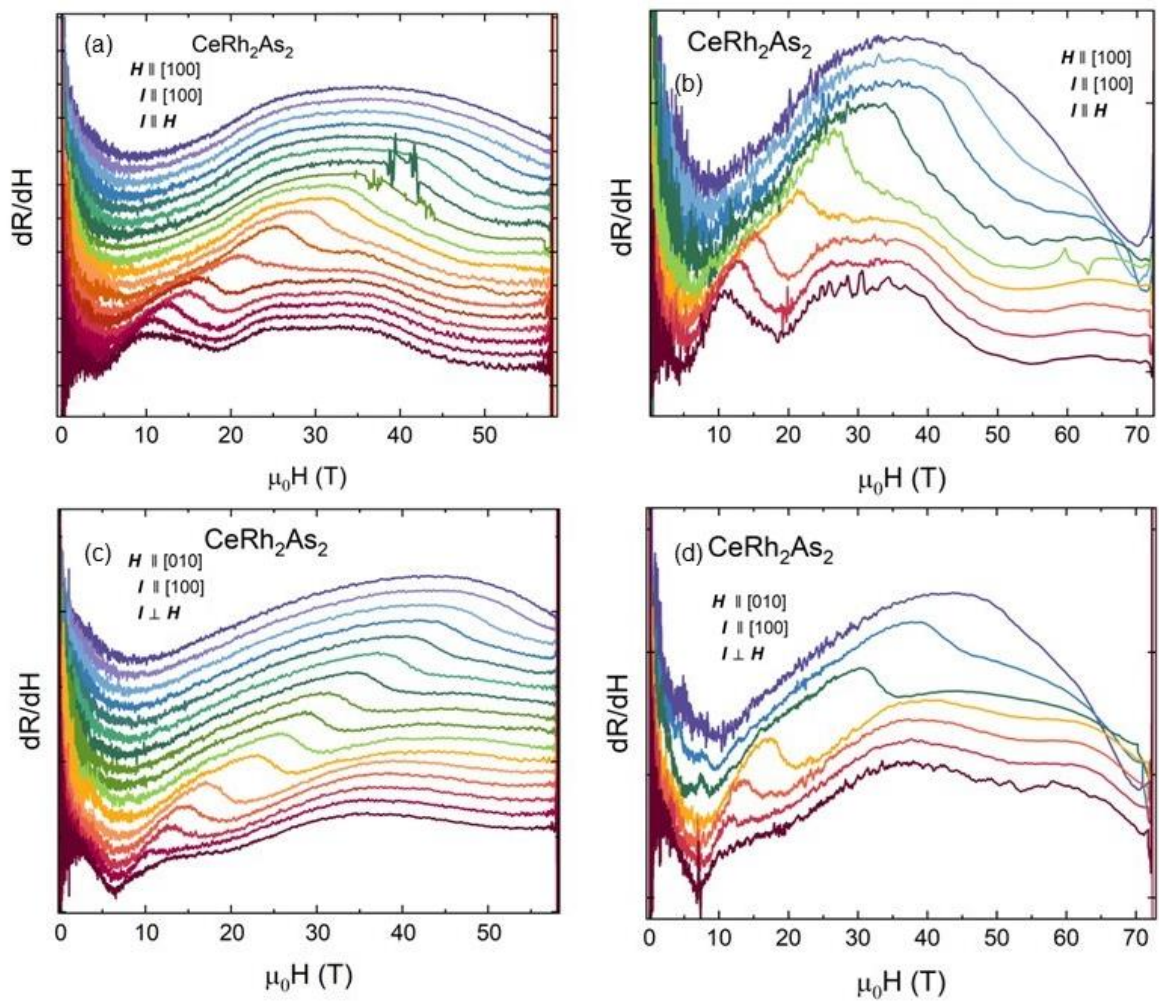

**Fig. S5.** The derivative dR/dH as function of magnetic field measured with current and field along [100] crystallographic direction.

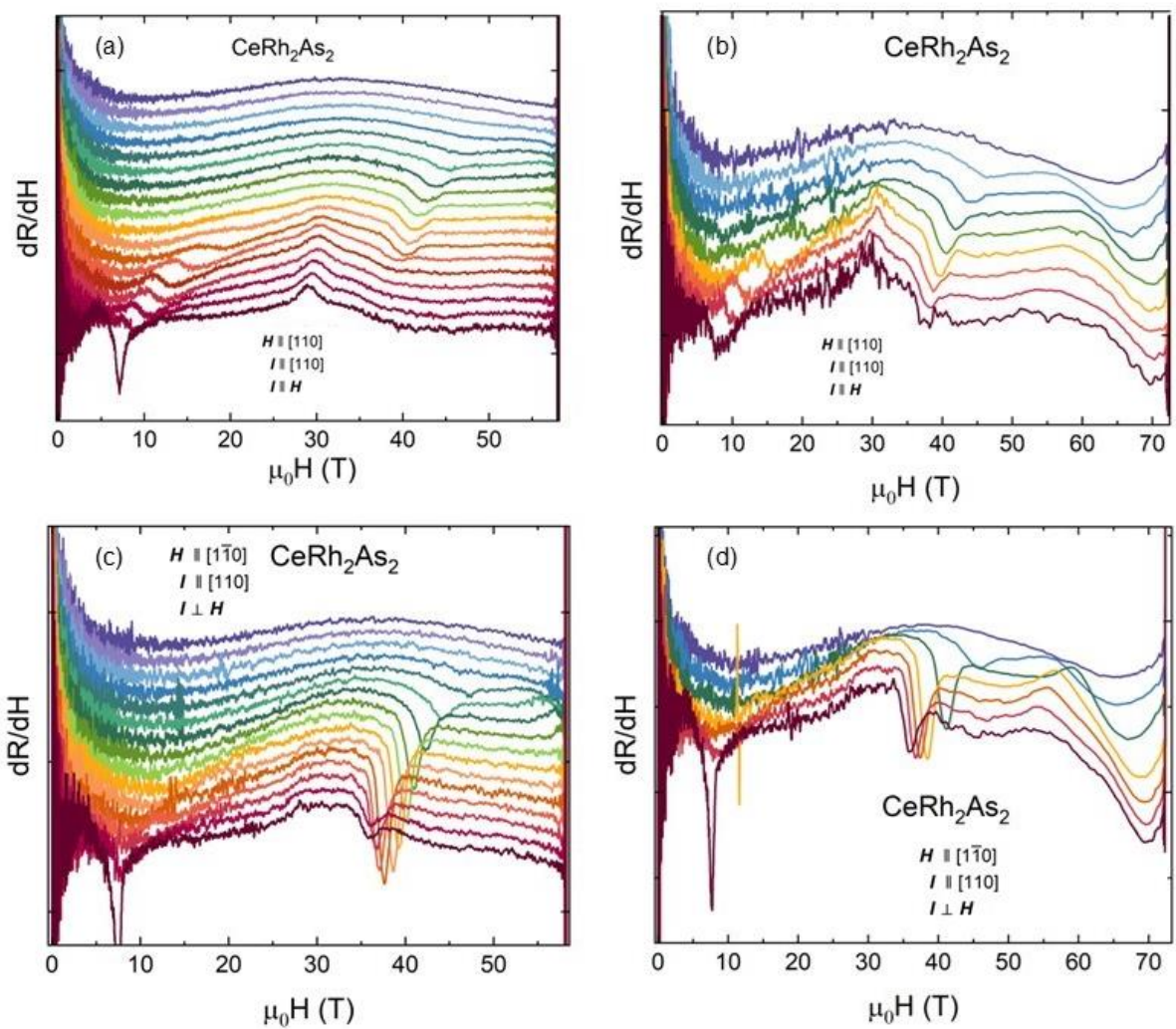

**Fig. S6.** The derivative dR/dH as function of magnetic field measured with current and field along [110] crystallographic direction.